\documentclass[aps,prb,psfig,twocolumn,showpacs]{revtex4-1}

\usepackage{graphicx}
\usepackage{color}
\usepackage{amsthm}
\usepackage{stmaryrd}
\usepackage{hyperref}

\newcommand{\mathsym}[1]{{}}

\begin{document}
\draft
\title{Current-induced spin polarization in graphene due to Rashba spin-orbit interaction}
\author{A. Dyrda\l$^{1}$,   J.~Barna\'s$^{1,2}$, and V. K.~Dugaev$^{3}$}
\address{$^1$Faculty of Physics, Adam Mickiewicz University,
ul. Umultowska 85, 61-614 Pozna\'n, Poland \\
$^2$  Institute of Molecular Physics, Polish Academy of Sciences,
ul. M. Smoluchowskiego 17, 60-179 Pozna\'n, Poland\\
$^3$Department of Physics, Rzesz\'ow University of Technology,
al. Powsta\'nc\'ow Warszawy 6, 35-959 Rzesz\'ow, Poland}
\date{\today }

\begin{abstract}
Spin polarization induced by an external electric field in graphene is considered theoretically in the linear response regime. The graphene is assumed to be deposited on a substrate which leads to the spin-orbit interaction of  Rashba type. The induced spin polarization is shown to be in the graphene plane and perpendicular to the electric field. However, the spin polarization changes sign when the Fermi level, whose position can be controlled by an external gate voltage,  crosses the Dirac points.
\end{abstract}

\maketitle

{\it Introduction} -- By mixing orbital and spin degrees of freedom, spin-orbit interaction leads to a variety of novel and interesting physical effects. Among them, there is a class of  transverse transport phenomena, where the transverse currents are driven by spin-orbit coupling. The most prominent example of these phenomena is the spin Hall effect (SHE), where an external electric field (or charge current due to this field) generates spin current flowing perpendicularly to the electric field. This phenomenon  is very attractive for possible  applications in spintronics devices, because it gives a unique possibility to create a pure spin current in nonmagnetic systems. The spin current, in turn, is a tool to control spin degree of freedom, and thus also orientation of the corresponding magnetic moments.

Long ago, Dyakonov~\cite{dyakonov71} predicted that an electric current flowing in a system with spin-orbit
(SO) interaction can induce not only perpendicular  spin current, but also  spin polarization of conduction electrons,
with the polarization vector perpendicular to the direction of current (and electric field). This phenomenon was studied later in a variety of systems exhibiting  spin-orbit interaction~\cite{edelstein90,aronov89,liu08,gorini08,wang09,schwab10}.
For example, Edelstein~\cite{edelstein90} has considered
spin polarization induced by electric field (current) in a two-dimensional electron gas with Rashba spin-orbit interaction, and found that the spin polarization is in the plane of the system and normal to the electric field. This follows from peculiar properties of electronic states in the Rashba field~\cite{rashba09}.
The current-induced spin polarization has been observed
experimentally, too~{\cite{kato04,silov04,sih05,yang06,stern06,koehl09,kuhlen12}.

In this letter we consider spin polarization induced by an electric field in  graphene, a strictly two-dimensional hexagonal lattice of carbon atoms. This crystal lattice can be considered as composed of two nonequivalent triangular sublattices.\cite{kane,castro}. Low energy electronic states in graphene are described by the relativistic Dirac model, with a characteristic conical energy spectrum. Fermi level of a free standing graphene consists of two nonequivalent Dirac points, $K$ and $K^\prime$. Intrinsic spin-orbit interaction opens an energy gap at the Fermi level (Dirac points), but this interaction in graphene is too small to generate measurable effects and may be neglected. Much stronger spin-orbit interaction can be induced by  a substrate. This interaction is known as the Rashba spin-orbit coupling. Its advantage over intrinsic spin-orbit interactions, in general, is due to the fact that the Rashba coupling parameter can be tuned by external electric field applied perpendicularly to the systems's plane, or simply by external gate voltage.

\begin{figure}[h]
\centering \includegraphics[width=1.0\columnwidth]{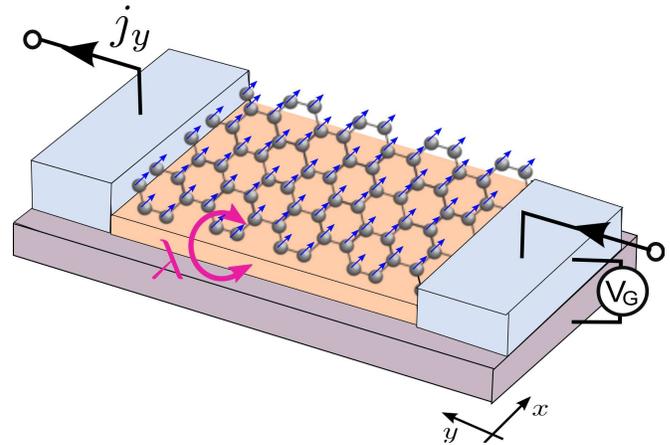}
  \caption{(color online) Scheme of the device based on graphene, where electric field (or charge current) flowing along the stripe (axis $y$) can induce spin polarization of carbon atoms. This spin polarization is  in the graphene plane and normal to the electric field. External gate voltage can be used to control position of the Fermi level in graphene. }
\end{figure}

It is already well known that the Rashba spin-orbit interaction in graphene gives rise to the spin Hall effect, with the spin current polarized perpendicularly to the graphene  plane and flowing in the direction normal to the electric field (charge current).\cite{sinova,dyrdal09} Here, we show that the Rashba coupling also leads to a current-induced spin polarization, which is oriented in the graphene plane and normal to the electric field, as shown schematically in Fig.1.

In section 2 we present the model we use in this paper and also  describe the method used to calculate the nonequilibrium spin polarization. Section 3 presents some numerical results on the current-induced spin polarization, while summary and final conclusions are in section 4.

{\it Model and method} -- We consider a system shown schematically in Fig.1. More specifically, we consider a strictly two-dimensional hexagonal lattice of  carbon atoms, with an external electric field applied in the graphene plane.
The low-energy electronic states of pristine graphene around the Dirac point $K$ are described by the Kane Hamiltonian $H_0^{K}$.\cite{kane} Since the intrinsic spin-orbit interaction in graphene is very small, as mentioned above, it will be neglected in the following considerations, and only the Rashba spin-orbit term induced by a substrate will be taken into account. Accordingly, the Hamiltonian $H_0^{K}$  for the $K$ point can be written in the following matrix form in the sublattice space:
\begin{eqnarray}
H^{K}_{0} = v \left(
      \begin{array}{cc}
        0 & k_{x} - i k_{y} \\
        k_{x} + i k_{y} & 0 \\
      \end{array}
    \right) \nonumber \\
    + \lambda \left(
                \begin{array}{cc}
                  0 & \sigma_{y} + i \sigma_{x} \\
                  \sigma_{y} - i \sigma_{x}  & 0 \\
                \end{array}
              \right) ,
\end{eqnarray}
where $k_{x(y)}$ are the wavevector components in the graphene plane, $v=\hbar v_F$ with $v_F$ denoting the electron Fermi velocity, $\lambda$ is the parameter of Rashba spin-orbit coupling, and $\sigma_{\alpha}$ ($\alpha =x,y,z$) are the Pauli matrices in the spin space. The parameter $v$ is connected with  the hoping integral $t$ between the nearest neighbors {\it via} the formula $v =ta\sqrt{3}/2$, where $a$ is the distance between nearest-neighbor carbon atoms. Without loosing generality we assume in the following that the parameter $\lambda$ is positive, $\lambda >0$, if not stated otherwise.

The spin polarization at $T=0$ can be calculated from the formula
\begin{equation}
S_{\alpha}(t) = - i \mathrm{Tr}\int \frac{d^{2} \mathbf{k}}{(2 \pi)^{2}} \Sigma_{\alpha} G_{\mathbf{k}}(t, t^\prime)|_{t^\prime =t+0},
\end{equation}
where $G_{\mathbf{k}}(t, t^\prime)$ is the zero-temperature Green function of the system in external field and $\Sigma_{\alpha}$ is the spin vertex function. In the case of graphene $\Sigma_{\alpha}$ may be written as
\begin{equation}
\Sigma_{\alpha} = \frac{\hbar}{2} I \otimes \sigma_{\alpha},
\end{equation}
where $I$ is a unit matrix in the pseudo-spin (sublattice) space. The relevant theoretical procedure includes now the Fourier transformation of the spin polarization and Green function with respect to the
time variables, and expansion of the Green's function in a series in external field. In the linear response theory only the first order term of the expansion is taken into account

\begin{figure}[h]
\centering \includegraphics[width=0.7\columnwidth]{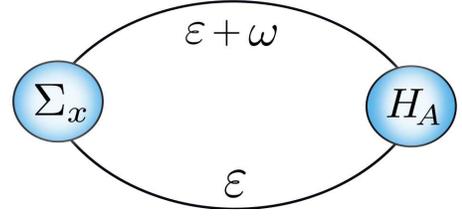}
  \caption{Diagram representing $x$ component of the current-induced spin polarization in the linear response regime. Here, $\Sigma_{x}$ is defined as  $\Sigma_{x} = \hbar I \otimes \sigma_{x}/2$, while $H_A$ is the perturbation due to external field, as described in the text. }
\end{figure}

To determine the current-induced spin polarization in the case under consideration, we assume that external electric field of frequency $\omega$ is along the axis $y$, and calculate the field-induced  spin polarization along the axis $x$ (see Fig.1). We have checked that the other components of spin polarization, i.e. those along the axes $y$ and $z$, vanish then exactly, at least in the linear response regime. Accordingly, the frequency-dependent $x$ component of the spin polarization induced by electric field is given in the linear response regime by the diagram shown in Fig.2. The corresponding analytical expression takes then the form (including contributions from both $K$ and $K^\prime$ points)
\begin{eqnarray}
S_{x} (\omega) = - i  {\rm{Tr}}\int\frac{d^{2}\mathbf{k}}{(2\pi)^{2}} \int\frac{d\varepsilon}{2 \pi}
\left(
      \begin{array}{cc}
      \sigma_{x} & 0\\
       0 & \sigma_{x} \end{array} \right)\nonumber \\
       \times\; G^0_{\mathbf{k}}(\varepsilon + \omega) H_{\mathbf{A}} G^0_{\mathbf{k}}(\varepsilon),
\end{eqnarray}
where $G^0_{\mathbf{k}}(\varepsilon)$ is the Green function corresponding to the Hamiltonian (1), and $H_{\mathbf{A}}$ is the perturbation due to the external  field, $H_{\mathbf{A}} = - e v_{y} A_{y}$.  Here, $A_{y}$ is the corresponding vector potential, which is connected to the electric field $E_{y}$ {\it via} the relation $A_{y} = - i E_{y}/\omega$, and $v_y$ is the $y$-component of the velocity operator. Equation (2) can be rewritten as
\begin{eqnarray}
S_{x} (\omega) = \frac{e E}{2\pi \omega} {\rm{Tr}}\int\frac{d^{2}\mathbf{k}}{(2\pi)^{2}} \int\frac{d\varepsilon}{2 \pi}
\left(
      \begin{array}{cc}
      \sigma_{x} & 0\\
       0 & \sigma_{x} \end{array} \right) \nonumber \\
         \times\;G^0_{\mathbf{k}}(\varepsilon + \omega) v_{y} G^0_{\mathbf{k}}(\varepsilon).
\end{eqnarray}
In the stationary limit, $\omega \rightarrow 0$, the above formula leads to the following expression:
\begin{eqnarray}
S_{x} = \frac{e}{2 \pi} E_{y} {\rm{Tr}}\int\frac{d^{2}\mathbf{k}}{(2\pi)^{2}}\left(
      \begin{array}{cc}
      \sigma_{x} & 0\\
       0 & \sigma_{x} \end{array} \right) G^{0R}_{\mathbf{k}}(\mu) v_{y} G^{0A}_{\mathbf{k}}(\mu),
\end{eqnarray}
where $G^{0R(A)}_{\mathbf{k}}(\mu)$ is the retarded (advanced) Green function corresponding to (1), and $\mu$ is the electrochemical potential measured from the energy corresponding to the Dirac points. The chemical potential can be controlled by an external gate voltage.
Taking into account the explicit form of the relevant Green functions
one finds
\begin{equation}
S_{x} = \frac{e}{(2 \pi)^{2}} E_{y} \int d k \frac{8 v \lambda \mu k (k^{4} v^{4} - \mu^{4} + 4 \mu^{2} \lambda^{2})}{\prod_{n = 1}^{4} (\mu - E_{n} + i \Gamma) (\mu - E_{n} - i \Gamma)},
\end{equation}
where $\Gamma = \hbar/2\tau$, with  $\tau$ denoting the momentum relaxation time.
The above equation presents a general formula for the current-induced spin polarization at $T=0$, from which one can find some simple analytical expressions valid in specific situations, as described in the following section.

{\it Analytical and numerical results} -- The integration over $k $ in Eq.(7) can be performed with the use of Cauchy's residue theorem, which
leads to simple analytical formulas for the spin polarization induced by electric field (current).
When $|\mu| < 2 \lambda$ one finds
\begin{equation}
S_{x} = \frac{e}{4 \pi} \frac{\mu (2 \lambda \pm \mu)}{v (\lambda \pm \mu)} E_{y} \tau ,
\end{equation}
where the upper and lower signs correspond to $\mu >0$ and $\mu <0$, respectively. In turn,  for  $|\mu| > 2\lambda$ we find
\begin{equation}
S_{x} = \pm \frac{e}{4 \pi} \frac{ 2 \mu^{2} \lambda}{v (\mu^{2} - \lambda^{2})} E_{y} \tau .
\end{equation}

\begin{figure}[h]
\centering
\includegraphics[width=1.0\columnwidth]{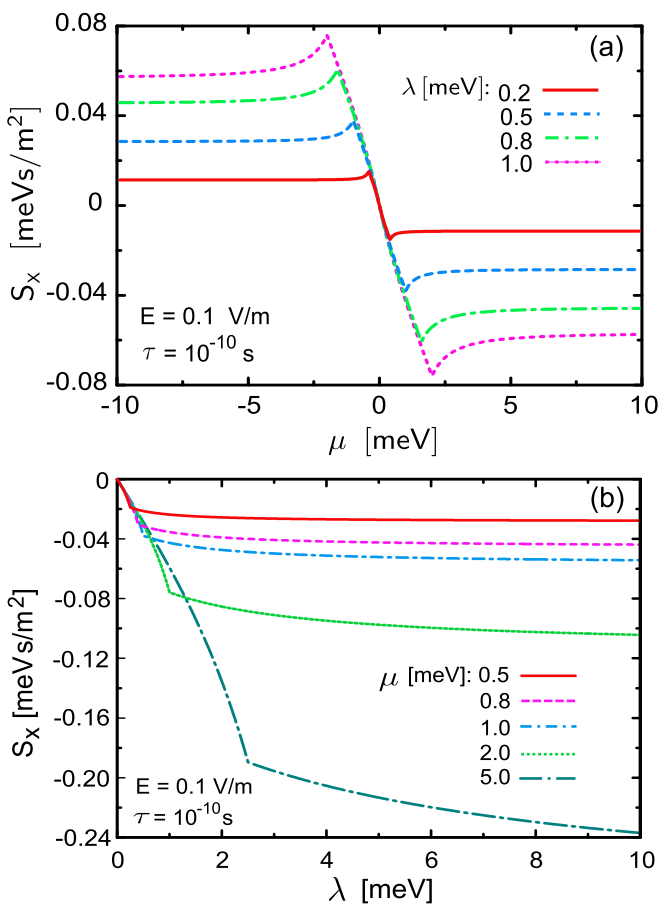}
  \caption{(color online) Spin polarization $S_x$ induced  by electric field, presented  as a function of the chemical potential $\mu$ for indicated values of the spin-orbit Rashba coupling parameter (a), and as a function of the Rashba parameter $\lambda$ for indicated values of the chemical potential. The other parameters are: $a = 1.4\AA$, $t = 2.9$eV, $\tau =10^{-10}$s, and $E_y=0.1$V/m. }
\end{figure}

The spin polarization is linear in electric field, which is obvious in the linear response regime. Apart from this, the spin polarization is also proportional to the relaxation time, similarly as in the case of spin polarization induced by electric field in a two-dimensional electron gas.  Variation of the spin polarization with the chemical potential $\mu$ is more complex and is shown in Fig.3(a) for specific values of the spin-orbit Rashba parameter. First, we note  that the spin polarization vanishes for $\mu=0$, ie. when the Fermi level is at the Dirac points. This is rather obvious due to a vanishing density of states at the Dirac points. When $\mu$ departs slightly from $\mu=0$, the absolute value of spin polarization increases. This increase is roughly linear in $|\mu|$ at small values of $\mu$. Indeed, in  the limit of $|\mu|<<\lambda$  one finds
\begin{equation}
S_{x} = \frac{e}{4 \pi} \frac{ 2 \mu  E_{y} \tau}{v} .
\end{equation}
Note, this rate of increase is linear in electric field $E_y$ and relaxation time $\tau$, but is independent of the Rashba parameter $\lambda$. The latter is clearly visible in Fig.3(a) for small values of $\mu$.

\begin{figure}[h]
\centering
\includegraphics[width=1.0\columnwidth]{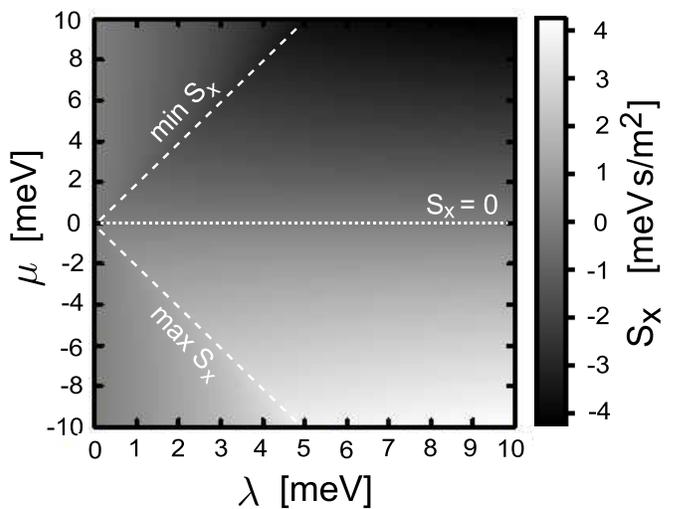}
  \caption{Spin polarization $S_x$ induced  by electric field, presented as a function of the chemical potential $\mu$ and Rashba spin-orbit coupling parameter  $\lambda$, calculated  for $a = 1.4\AA$, $t = 2.9$eV,  $\tau =10^{-10}$s, and $E_y=0.1$V/m. Local maxima in $|S_x|$ as a function of $\mu$ are marked by dashed lines. }
\end{figure}

When $|\mu|$ increases further,
$|S_x|$ reaches a maximum at some point (which depends on the coupling parameter $\lambda$) and then weakly decreases with a further increase in $|\mu|$. Finally, $S_x$ saturates at large values of $|\mu|$,
\begin{equation}
S_{x} = \pm \frac{e}{4 \pi} \frac{ 2 \lambda E_{y} \tau}{v}
\end{equation}
for $\mu^2 >>\lambda^2$.

The above described behavior of spin polarization and the formulas for the two limiting situations, as well as the data presented in Fig.3(a) clearly show that the spin polarization is negative for $\mu>0$ and positive for $\mu<0$ (note that $e$ is the electron charge). Thus, the current-induced spin polarization in graphene changes sign at $\mu=0$. This behavior is different from that found  in a two-dimensional gas, where spin polarization does not change sign with the chemical potential.\cite{dyrdal13} Note, that the spin polarization would also change sign when the spin-orbit parameter $\lambda$ would be negative.

Variation of spin polarization with the Rashba parameter $\lambda$ is shown in Fig.3(b) for indicated values of the chemical potential $\mu$. When $\lambda>>\mu$, the spin polarization saturates and becomes independent of $\lambda$ in agreement with Eq.(10). In turn, for $\lambda <<\mu$, the absolute value of spin polarization grows linearly with $\lambda$, as follows from Eq.(11).

Figure 4 presents density plots for the current-induced spin polarization as a function of the chemical potential $\mu$ and Rashba coupling parameter $\lambda$.
The curves presented in Fig.3 are cross sections of this figure taken either at constant $\lambda$ (Fig.3(a)) or at constant $\mu$ (Fig.3(b)). Maxima in the absolute value of the spin polarization are clearly seen in this figure. These maxima appear along two lines (the dashed lines in Fig.4) in the plane determined by chemical potential $\mu$ and spin-orbit parameter $\lambda$ -- one for positive $\mu$ and one for negative $\mu$.

{\it Summary} -- We have analyzed the spin polarization in graphene induced by an external electric field (or associated current). To find the stationary nonequilibrium polarization in the linear response we have used the Green function method. From this, we have derived some analytical formulas for the induced spin polarization. The spin polarization is shown to be in the graphene plane and normal to the current orientation. Sign of the spin polarization depends on the chemical potential $\mu$ and changes at $\mu=0$. Moreover, sign of the polarization also depends on the sign of the spin-orbit Rashba parameter $\lambda$.

When absolute value of the chemical potential, $|\mu |$, is much larger than the Rashba parameter, the spin polarization grows linearly with $\lambda$ and is roughly independent of $\mu$. In turn, when $|\mu |$ is much smaller than the Rashba parameter, the spin polarization is independent of $\lambda$, but varies linearly with the chemical potential $\mu$. Between these two limiting situations, amplitude of the spin polarization reaches a maximum at some chemical potential. Position of these points changes linearly with the  Rashba parameter.

Since the spin polarization is induced by current flowing due to an external electric field, one can expect a similar spin polarization as a result of current flowing due to temperature gradient instead of electric field. Indeed, such a spin polarization induced by thermal currents was proposed recently in the case of two-dimensional electron gas.\cite{wang10,dyrdal13} Such a thermally induced spin polarization should also appear in graphene. This problem, however, requires a detailed analysis and will be considered in a separate paper.

{\it Acknowledgments} --  This work has been supported  by funds of the National Science Center in Poland through Grants No. DEC-2011/03/N/ST3/02353 as well as DEC-2012/04/A/ST3/00372 and DEC-2012/06/M/ST3/00042.  A.D. also acknowledges support by the
Adam Mickiewicz University Foundation.


\end{document}